\documentclass[12pt]{iopart} 
\usepackage{graphicx}
\begin{document}

\title[The unconventional two-parameter quantum valley pumping in graphene with a topological line defect]
{The unconventional two-parameter quantum valley pumping in graphene with a topological line defect}

\author{C D Ren$^{1}$, L Cui$^2$, W T Lu$^3$, H Y Tian$^{3}$, S K Wang$^4$}

\address{$^1$ Department of Physics, Zunyi Normal College, Zunyi 563002, China}
\address{$^2$ Suqian university, Suqian 223800, China}
\address{$^3$ School of Physics and Electronic Engineering, Linyi University, Linyi 276005, China}
\address{$^4$ College of Science, Jinling Institute of Technology, Nanjing 211169, China}

\vspace{10pt}

\begin{abstract}
Based on the Keldysh Green's function method, we report an unconventional two-parameter quantum pumping in graphene with a line defect. It is found that different from the conventional sinusoidal relation, the pumped current in this device is cosinusoid dependence on the phase difference $\varphi$ between the two pumping potentials, which adopts its positive/nagative maximum value at $\varphi=0/\pi$ while tends to zero at $\varphi=\pi/2$. This phenomenon is related to the peculiar valley tunneling characteristics across the line defects and the exchange of valley indices on both sides of the line defect. Moreover, the pumped currents from the two valleys will flow in opposite directions along the line defect, indicating that the controllable valley current can be pumped out in the line defect without the application of strain field in graphene.
\end{abstract}

\vspace{2pc}
\noindent{\it Keywords}: Two-parameter quantum pumping, Valley current, Line defect \\
\maketitle

\section{Introduction}
In recent years, a newly emerged electronics, the valleytronics, has attracted widespread attention with the successful preparation of the two-dimensional(2D) hexagonal crystals, including graphene, silicene, group-VI transition-metal dichalcogenides (TMDs), etc\cite{1,2,3,4,5,6,7,8,9,10}. The concept of valley originates from two degenerate but nonequivalent
energy bands at the local minimum in the conduction band or local maximum
in the valence band in these materials\cite{11,12}. The two valleys, $K$ and $K'$, are related by time-reversal symmetry and have opposite phase factors in the wave function\cite{13}.
Superior to the charge and spin degree of freedom(DOF), the valley DOF have
greater advantages in storing and processing information.
Therefore, the major task in this field is to filter or seperate the valley states to achieve an effective valley currents.

A parametric quantum pumping is an effective medium to generate a pure charge or spin currents by using periodic variation of physical parameters such as ac gate potential between unbiased leads\cite{14}. The broken of spatial inversion symmetry(SIS) is the necessary condition to pump a pure electronic current. For the nature SIS breaking system, the currents can be pumped out with only a single ac gate potential\cite{15,16,17}. More commonly, a quantum pumping involves two or more ac gate potentials with a definite phase difference to break the SIS of the system\cite{18,19,20}.
The dc pumped current is sinusoidal dependence on the
phase difference $\varphi$ between the pumping potentials and
adopts its maximum(minimum) value as $\varphi=\pi/2(0)$.
The dc currents in this phenomenon comes from the interfere of the electronic waves due to the phase difference between ac gate voltages\cite{18}. The theoretical approach was proposed by Thouless\cite{19} and realized in a quantum dot\cite{20}. In graphene, the charge and spin currents have been extensively studied in various nanoscale systems with multi-parameter devices in the adiabatic or nonadiabatic quantum pumping regime\cite{21,22,23,24,25,26}. The quantum pumping can be considered as the adiabatic regime when the period of oscillating voltages is much greater than transmission time of carriers throughout the system, while a finite pumping frequency corresponds to the nonadiabatic regime\cite{23}. By contrast, the multi-parameter quantum valley pumping devices were fewly studied in graphene and the current works are depending on the strain field in a graphene sheet in an adiabatic regime. For instance, Jiang $et\ al$ pointed out that the bulk valley current can be generated in graphene by periodic modulation of strain field and chemical potential\cite{27}. Wang $et\ al$ proposed a three potential barriers junction including two regions of strained graphene with antiparallel magnetizations to pump the pure valley current in graphene\cite{28}. Yu $et\ al$ investigated the pumped valley currents in zigzag graphene nanoribbons with multi-pumping potentials\cite{29}. The pumped valley current can also be generated
in graphene by introducing a series of antiparallel magnetic fields and the
electrical potential barriers\cite{30}. However, these valley currents are not easily implemented experimentally.

For practical applications, the large-area graphene grown by chemical
vapor deposition technique are always found to be polycrystalline in nature\cite{31,32,33,34},
consisting of two domains with different crystallographic orientations and the grain boundaries(GBs) stitching them together. In valleytronics, more attention has been paid to a special kind of polycrystalline called the line defect\cite{34,35,36,37} which is composed of
carbon atom pentagons and octagons since the line defect valley filter has already been observed experimentally on a metallic substrate\cite{34}. Distinct from its pristince form, the
lattice vectors are inversed on both sides of the line defect and the
two $A/B$ sublattices as well as the valley index are also swaped. Namely, the $K(K')$ valley on one domain is equivent to the $K'(K)$ valley on the other domain, therefore, the wave function of a given valley state should be hermione conjugated on both sides of the line defect.
The transmission of the valley states across the line defect are angle dependent,
only one valley state can pass through the line defect at large incident angle
and induces nearly $100\%$ valley polarization.
In the line defect, the quasi-one-dimensional electronic states are
distributed around the line defect\cite{38} which palys a key role in the
quantum parametric pumping. Recently,  Ren\cite{15} found that when a single-parameter nonadiabatic pumping is applied on the line defect, the pumped currents from the two valleys
will flow in opposite directions along the line defect due to the asymmetric
scattering in the photon-assisted processes.
Nevertheless, the valley pumping mechanism for the two-parameter nonadiabatic pumping device in the line defect is still unclear which needs to be clarified.

In this paper, based on the Keldysh Green's function method, we theoretically study the quantum valley pumping phenomenon in the line defect using two ac fields with a definite phase difference $\varphi$ in the nonadiabatic regime. It is found firstly that the same as the single parameter pumping, the pumped valley current is also scattering angle $\alpha$ dependent where the pumped current from $K(K')$ valley only occurs at about $\alpha\approx-\pi/2(\pi/2)$ when the two ac fields are applied on each side of the line defect. Secondly, and also the most importantly, the pumped current is cosinusoid dependence on $\varphi$ rather than the sinusoidal relation as in the conventional devices. The maximum valley current occurs at $\varphi=0$ or $\pi$ while it equals about zero at $\varphi=\pi/2$. 
This is because the wave function of a given valley is hermione conjugated on both sides of the line defect, 
the valley state from only one electrode can enter the center scattering region at a certain scattering angle $\alpha$ and form a pumped current as $\varphi=0$.
Therefore, the pumped current
from the $K$ valley can even appear at $\alpha=\pi/2$ when the two ac fields are applied on the right side of the line defect.
We provide a pure electrical means of regulating the valley current without introducing the strain field, and also reveal the unconventional pumping mechanisms present in the line defect.

\section{Model}
\begin{figure*}
	\includegraphics[width=4.5in]{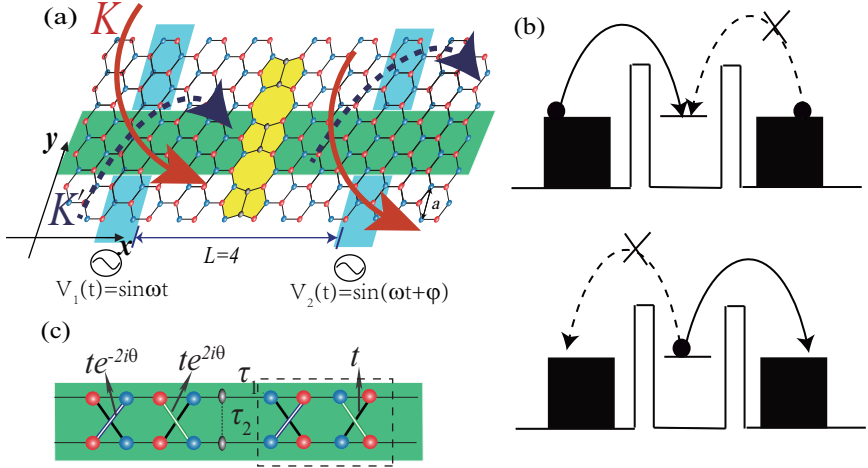}
	\caption{(Color online) (a)
		Schematic diagram of the two-parameter valley pumping, where the two ac-fields with a definite phase difference $\varphi$ are applied in the gray region.
		The red real line(blue solid line) arrow represents the flowing directions
		of $K(K')$ valley state and $L$ is the distance between the two pumping potentials
		in units of $\sqrt{3}a$ with $a(a=0.246$nm) the lattice constant of graphene.
		(b) Schematic representation for the unidirectional transmission process in the two-parameter pumping as $\varphi=0$.
		(c)The simplified lattice model of the line defect where $\theta=k_ya$ is from the Fourier transformation along the $y$ axis and the dashed rectangle defines a
		supercell of pristine graphene.}
	\label{fig:1}
\end{figure*}
The two-parameter pumping device we considerd here is shown in Fig. 1(a).
An infinite line defect sample is lying in the $x-y$ plane and two time dependent pumping potentials(ac fields) are applied in the gray region.
The time-dependent Hamiltonian written in the second
quantization notation is\cite{15}
\begin{eqnarray}
	H_0=\sum_{i}\epsilon_ic_{i}^{\dag}c_{i}-t\sum_{\langle i,j\rangle}c_{i}^{\dag}c_{j}
	-t_2\sum_{\langle\gamma\delta\rangle}c_{i_y,\gamma}^{\dag}c_{i_y,\delta}-
	t_1\sum_{\langle i,\gamma\rangle}c_{i}^{\dag}c_{i_y,\gamma}+
	\sum_{i}V_{ac}\sin(\omega t+\varphi)c_{i}^{\dag}c_{i}+c.c..
\end{eqnarray}
The first term denotes the on-site energy, the second to fourth
terms represent different nearest neighbour couplings with
hopping energies $t, t_1$ and $t_2$, as shown in Fig. 1(c).  $c_{i}^{\dag}$($c_{i_y,\gamma/\delta}^{\dag}$)
is the quasi-particle creation operator at site $i$(the defect atoms).
In the line defect of graphene, the variations of the bond lengths around the
line defect varies a little compared with the defect-free region, therefore
it is reasonable to set $t_{1}\approx t_{2}\approx t$ where we can set
$t=1$(for which $t\approx$3.1 eV) as the energy unit.
The last term is the ac-field applied in the scattering region
with the magnitude $V_{ac}$ and  frequency $\omega$. $\varphi$ is the
phase difference between the two pumped sources.

In the nonadiabatic regime, the Keldysh Green's function method is an effective technique
to calculate the pumped currents in a pumping cycle $\tau$\cite{15,21}, which can be wirtten as ($e=\hbar=1$),
\begin{eqnarray}
	I_\mu=-\frac{1}{\tau}\int_{0}^{\tau}\int dt_1\mathrm{Tr}[G^r(t,t_1)\Sigma_\mu^<(t_1,t)+G^<(t,t_1)\Sigma_{\mu}^a(t_1,t)+h.c.],
\end{eqnarray}
$\mu$ denotes the electrode($\mu=L,R$) and then the above formula corresponds to the current
flowing into the $\mu$ electrode in a pumping cycle.
Here, $G^{r,<,a}(t,t_1)$ are the retarded, less and advanced Green's functions in the
scattering region.
\begin{eqnarray}
	G_{ij}^{r(a)}=\mp \mathrm{i}\phi(\pm t\mp t')\langle[c_i(t),c_j^\dag(t')]\rangle,\\
	G_{ij}^{<}=i\langle c_i^\dag(t)c_j(t')\rangle, \nonumber
\end{eqnarray}
where$\langle\cdots\rangle$ is the quantum statistic average and $\phi(\pm t\mp t')$ is a step function. $\Sigma_{\mu}^{r}$ is the retarded self energy of the semi-infinite
electrode which can be evaluated in terms of the
isolated resevoir surface Green's function $g_{00}$ and the coupling matrix $H_{S,\mu}$,
$\Sigma_{\mu}^{r}=H_{\mu,S}g_{00}H_{S,\mu}$. $\Sigma_{\mu}^{a}=[\Sigma_{\mu}^{r}]^\dag$
and $\Sigma_{\mu}^{<}=(\Sigma_{\mu}^{a}-\Sigma_{\mu}^{r})f_\mu$
with $f_\mu=[e^{E/k_BT}+1]^{-1}$ the Fermi function for the lead $\mu$. In the calculations,
we set $f_L=f_R=f$ because no bias is applied on the system.

We also explore the perturbation method to evaluate $G^{r,a,<}$
and taking the pumping potentials as the perturbed term.
According to the Dyson equation, the Green's function in Keldysh space $G^k$ can be given by:
\begin{eqnarray}
	G^{k}(t,t')=G^{k0}(t,t')+\int dt_1G^{k0}(t,t_1)V^k(t_1)G^{k0}(t,t')+...,
\end{eqnarray}
where $G^{k}(t,t')$ is defined as
\begin{eqnarray}
	G^{k}(t,t')=
	\left(
	\begin{array}{cc}
		G^t(t,t') & G^<(t,t') \\
		G^>(t,t') & G^{\bar{t}}(t,t') \\
	\end{array}
	\right),
\end{eqnarray}
with
\begin{eqnarray}
	G^{t}(t,t')=-\mathrm{i}\phi(t-t')\langle c(t)c^\dag(t')\rangle+\mathrm{i}\phi(t'-t)\langle c(t')^\dag c(t)\rangle, \nonumber \\
	G^>(t,t')=-\mathrm{i}\langle c(t)c^\dag(t')\rangle,\\
	G^{\bar{t}}(t,t')=-\mathrm{i}\phi(t'-t)\langle c(t)c^\dag(t')\rangle+\mathrm{i}\phi(t-t')\langle c(t')^\dag c(t)\rangle. \nonumber
\end{eqnarray}
The above formulae are the time-order, greater, and anti-time-order Green's function.
The perturbation potential $V^k$ in the
Dyson equation can be defined in the Keldysh space as
\begin{eqnarray}
	V^{k}(t)=
	\left(
	\begin{array}{cc}
		H_1 & 0 \\
		0 & H_1 \\
	\end{array}
	\right).
\end{eqnarray}
The four component Green's functions of $G^k$ have the following relations:
\begin{eqnarray}
	G^t=G^<+G^r,   \nonumber\\
	G^{\bar{t}}=G^a-G^a,\\
	G^>=G^t-G^a.  \nonumber
\end{eqnarray}
For the unperturbed term, the Green's function $G^{k(r)0}$ can be evaluated
according to $G^{r0}(E)=1/(E+\mathrm{i0^+}-H)$, where $H$ includes
the Hamiltonian of the scattering region and the self-energy
from the two leads.
The unperturbed lesser Green’s function is
obtained by $G^{<0}(E)=[G^{a0}(E)-G^{r0}(E)]f$, with $f$ being the
Fermi distribution function. Therefore, $G^{k0}$ can be calculated
according to Eqs. (7) and (8).
The bilinear response of the pumped current in lead $\mu$ can be obtained
with the perturbation method according to the above formulas:
\begin{eqnarray}
	I_\mu=-\mathrm{i}\sum_{i,j\in S}\frac{V_{ac}^2}{4}\int d\epsilon\mathrm{Tr}[\Gamma^\mu(\epsilon)X^{i,j}_{m_2,m_1}(\epsilon)]\times  \nonumber\\
	\{[f(\epsilon-\omega)-f(\epsilon)][G_{ij}^{r0}(\epsilon-\omega)-G_{ij}^{a0}(\epsilon-\omega)]e^{\mathrm{i}\varphi}\\+
	[f(\epsilon+\omega)-f(\epsilon)][G_{ij}^{r0}(\epsilon+\omega)-G_{ij}^{a0}(\epsilon+\omega)]e^{-\mathrm{i}\varphi}\}\nonumber, \\
	X^{i,j}_{m_2,m_1}(\epsilon)=G_{m_2,i}^{r0}(\epsilon)G_{j,m_1}^{a0}(\epsilon)  \nonumber
\end{eqnarray}
where $\Gamma=\mathrm{i}[\Sigma^r-\Sigma^a]$ is the line-width
function, $S$ denotes the scattering region and
$m_{1(2)}$ indicates the positions where the scattering region is contacted with the leads. $\varphi$ is the phase difference between the two ac pumping potentials.

The line defect is infinite along the $y$ direction, as shown in Fig. 1(a), where the green region represents a unit cell. Therefore, $k_y$ is a conserved quantity and 
the creation(annihilation) operator
can be transformed in the momentum space according to the Fourier transformation:
\begin{eqnarray}
	c_{i}^{\dag}=\sum_{k_y}c_{k_y,i_x}e^{-2ik_yi_ya},c_{i}=\sum_{k_y}c_{k_y,i_x}e^{2ik_yi_ya},  \nonumber \\
	c_{i_y,\gamma}^{\dag}=\sum_{k_y,\gamma}c^{\dag}_{k_y,\gamma}e^{-2ik_yi_ya},c_{i_y,\gamma}=\sum_{k_y,\gamma}c_{k_y,\gamma}e^{2ik_yi_ya}.
\end{eqnarray}
The Hamiltonian matrix of a supercell can be depicted in the following form:
\begin{eqnarray}
	H_{k_y}=-\sum_{i}\varphi_{i,1}^{\dag}\hat{T}_{1}\varphi_{i,2}-\sum_{i}\varphi_{i,2}^{\dag}\hat{T}_{2}\varphi_{i,3} \nonumber \\
	-\sum_{i}\varphi_{i,3}^{\dag}\hat{T}_{1}^{\dag}\varphi_{i,4}-\sum_{i\neq-1}\varphi_{i,4}^{\dag}\hat{T}_{2}^{\dag}\varphi_{i+\hat{x},1} \\
	-\varphi_{\bar{1},4}^{\dag}\hat{T}_{2}\varphi_{0}-\varphi_{0}^{\dag}\hat{T}_{2}\varphi_{1,1}-\varphi_{0}^{\dag}\hat{T}_{3}\varphi_{0}
	+h.c., \nonumber \\
	\hat{T}_{1}=\left(
	\begin{array}{ccc}
		1 & 1\\
		e^{-i2\theta} & 1\\
	\end{array}
	\right),
	\hat{T}_{2}=\left(
	\begin{array}{ccc}
		1 & 0\\
		0 & 1\\
	\end{array}
	\right),
	\hat{T}_{3}=\left(
	\begin{array}{ccc}
		0 & 1\\
		1 & 0\\
	\end{array}
	\right).  \nonumber
\end{eqnarray}
Here, $\varphi _{i,\gamma}^{\dag}=\left[ c_{{{k}_{y}},i,\gamma,1}^{\dag}, c_{{{k}_{y}},i,\gamma,2}^{\dag}\right]$$(\bar{i}=-i)$,
$\hat{x}$ represents the
unit length between the neighboring supercells in the graphene
part, $i$ is the position of a supercell, $\gamma$ takes an
integer number from 1 to 4 denoting the different columns in
a supercell, and $1/2$ in $ c_{{{k}_{y}},i,\gamma,1/2}^{\dag}$ corresponds to the up/down
site in the same column in Fig. 1(c).

In the line defect, the two Dirac points are located at $[0,\pm\pi/3a]$.
Therefore, an electron in two valleys satisfies
the following relations: $k_x=q_x$ and $k_y=q_y\pm\pi/3a$ where $q_x$ ($q_y$)
represents the group velocity of electrons along the $x$ ($y$) direction.
Combining this with the linear dispersion relation of the Dirac electrons $E=\frac{\sqrt{3}q}{2}=\frac{\sqrt{3}\sqrt{q_x^2+q_y^2}}{2}$,
the pumped current from $\eta(\eta=K,K')$ valley in lead $\mu$
can be calculated as a function of the electron scattering angle $\alpha$($\alpha=\arctan(q_y/q_x)$)
according to the Eq. (9). In fact, $\alpha$ is also the phase factor in the wave
function of the valley states\cite{13}. For graphene, this phase
factor is opposite for different valleys, however, it is also opposite
for the same valley on both sides of the line defect because the valley indices
are exchanged now.

\section{Numerical Results}
In this section, we will numerically investigate the novel quantum pumping mechanism
in the line defect with two ac fields.
In the calculations, we set the length between the two ac fields is $L=20$ and
$V_{ac}=0.005$ to justify the perturbation theory.

\begin{figure}
	\includegraphics[width=4.0in]{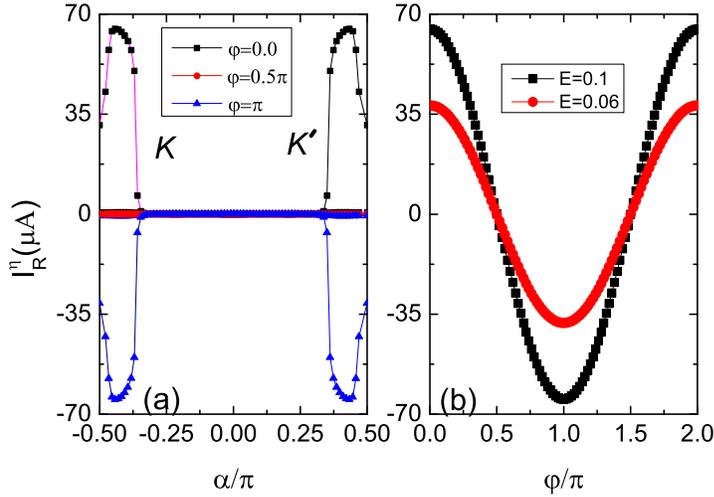}
	\caption{(Color online) (a)The pumped valley currents flowing into the right lead $I_R^\eta$ as a function of $\alpha$ for different phase difference $\varphi$.(b) $I_R^K$ as a function of $\varphi$ for different Fermi energies at $\alpha=-0.44\pi$. The frequency is $\omega=0.01$.}
	\label{fig:2}
\end{figure}
In Fig. 2(a), the pumped currents of the two valleys flowing into the right lead($\mu=R$) are plotted as a function of the scattering angle $\alpha$ for different phase difference $\varphi$. Similar to the single quantum pumping device, the pumped currents are also angle dependent. For instance, $I_R^K$ only occurs at the angle interval $\alpha\in[-0.5\pi,-0.38\pi]$ while $I_R^{K'}$ occurs at the opposite angle interval $\alpha\in[0.38\pi,0.5\pi]$.
However, it is interestingly found that the pumped valley current has a large
magnitude as $\varphi=0(\pi)$ while it even equals zero as $\varphi=\pi/2$, which is totally different from the conventional pumping phenomenon. In Fig. 2(b), 
we further plot $I_R^K$ as a function of $\varphi$ for different energies.
It is shown that $I_R^{K}\sim\cos\varphi$
rather than $\sin\varphi$ as in the traditional devices.
This phenomenon can be explained with a simple one-way transmission 
mechanism of the valley state, as depicted in Fig. 1(b). For the 
traditional two parameter pumping devices, the carriers in the left and right electrode can not simultaneously absorb photons and enter into the center
region as $\varphi=0$ due to the pauli exclusion principle. However, this
restriction is relieved in the line defect now. For instance, a $K$ valley carrier can absorb a photon and enter the center region
from the left lead at the angle $\alpha\approx-\pi/2$ while the other
transmission process is prohibitted, because the wave function of right $K$ valley carrier is hermione conjugated with the left one. 
Or, the $K$ valley in the right side of line defect has the same transmission properties with the $K'$ valley in the left side of the line defect. It is noted that the valley currents flowing into the left and right leads are identical, $I_L^{\eta}=I_R^{\eta}$.

\begin{figure}
	\includegraphics[width=4.0in]{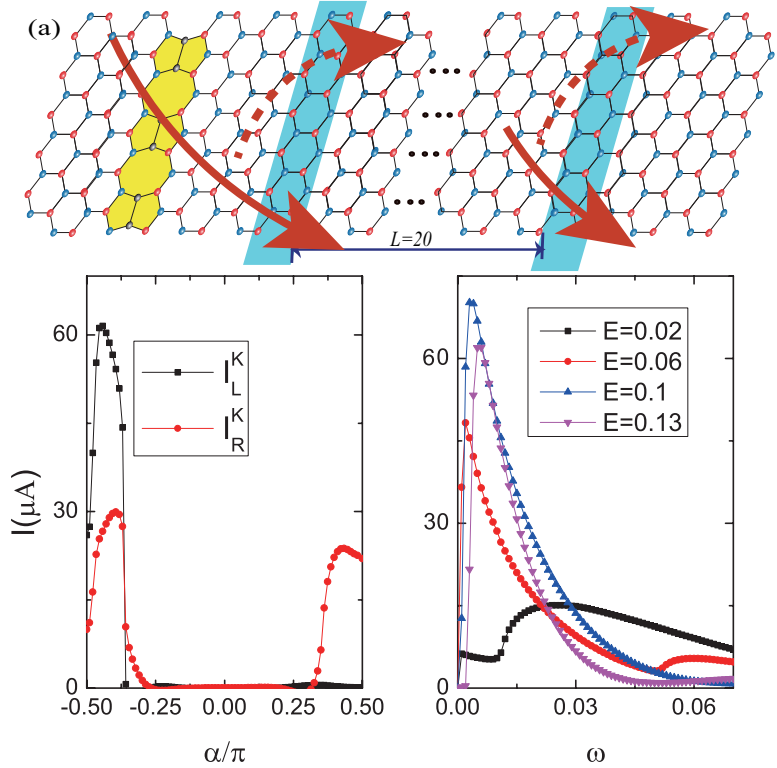}
	\caption{(Color online)  (a)Schematic diagram of the valley pumping phenomenon when
		the two pumping potentials are applied on the right side of the line defect, and
		(b) the corresponding valley currents $I_{L/R}^K$ versus $\alpha$ as $\omega=0.01$. (c) $I_{L}^K$ as a function of $\omega$ for different Fermi energies at $\alpha=-0.44\pi$. The distance between the line defect and the left ac field is set as $L=2$.}
	\label{fig:3}
\end{figure}

To further authenticate this viewpoint, we also investigate the pumped valley currents
when the two ac fields are applied on the right side of the line defect, as shown in Fig. 3.
Interestingly, it is found that the pumped current flowing into the left lead($I_L^K$) changes
a little while $I_R^K$ varies drastically, the current intensity becomes about the half of the original one in the angle region $\alpha\in[-0.5\pi,-0.38\pi]$ while the remaining
current appears at $\alpha\in[0.38\pi,0.5\pi]$. Obviously, the pumped current at $\alpha\in[0.38\pi,0.5\pi]$ is contributed by the $K$ valley carriers from the right
side of the line defect, as shown Fig. 3(a) where the real line arrow/solid line arrow denote the pumping process from the left/right side of the line defect. 
Therefore, it is not a good choice to set the two ac fields on the same side of the line defect
which will reduce the valley polarization.
Moreover, we also plot $I_L^K$ as a function of $\omega$ for different energies. It can be seen that the maximum current appears at about $E=0.1$ with low $\omega(\approx0.01)$. This is related to the distribution of the electron density of states(DOS) in the line defect, which adopts its maximum value at about $E=0.11$\cite{39} while it decreases as the Fermi energy departs from this point and even approaches zero at the Dirac point. Therefore, one can obtain a large pumped valley current when the
the two pumping potentials are applied on both sides of the line defect by tunning the Fermi energy
at about $E=0.1$.

\section{Conclusion}
In summary, we have investigated a kind of unconventional two-parameter nonadiabatic quantum pumping in graphene with a line defect. It is found that due to the peculiar transmission feature of the valley state across the line defect and the lattice symmetry relationship, the pumped current is cosinusoid with respect to the phase difference between the two pumping potentials, rather than the traditional sinusoidal relation. Moreover, the pumped valley currents are also angle dependent, the current from $K$ valley flows along one direction while that from $K'$ valley flows along the opposite direction. Currently, the valley pumping in graphene are all dependent on the strain field while the strain tuning is still unavailable up to now due to the complex mechanical setup.
It is hoped this scheme can have useful
applications in the development of graphene valleytronic
devices.

\end{document}